\documentclass[12pt]{article}
\usepackage{pic04}
\usepackage{hyperref}
\usepackage{url}
\usepackage{graphicx}

\begin{document}
\title{\bf ATLAS RPC Cosmic Ray Teststand at INFN Lecce}
\author{
G. Cataldi, G.Chiodini, R. Assiro, P. Creti, G. Fiore, F. Grancagnolo, A. Miccoli,\\
R. Perrino, S. Podkladkin, M. Primavera, and A. Ventura\\
{\em INFN - via Arnesano 73100, Lecce - Italy}\\
M. Bianco, E. Brambilla, C. Chiri, M.R. Coluccia, R. Gerardi, E. Gorini, \\
S. Spagnolo, and G. Tassielli.\\
{\em Dipartimento di Fisica - via Arnesano 73100, Lecce - Italy}\\
}
\maketitle

\baselineskip=14.5pt
\begin{abstract}
We describe the design and functionality of the cosmic ray teststand 
built at INFN Lecce for ATLAS RPC quality control assurance.\\
\end{abstract}

\baselineskip=17pt

\section{Introduction}
\begin{figure}[htbp]
  \centerline{\hbox{\includegraphics[width=10cm]{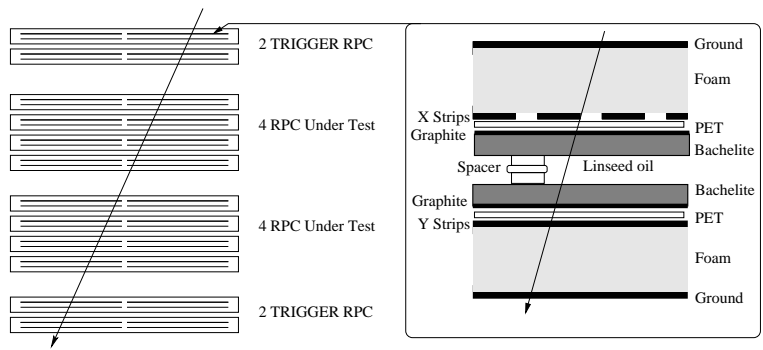}}
  }
 \caption{\it Cosmic ray teststand at INFN Lecce (left).
              Single gap ATLAS RPC (right). ATLAS RPC units consist of two layers of one or two adjacent single gap RPC.
	      \label{fig1.eps} }
\end{figure}

Resistive Plate Chamber ( RPC \cite{Santonico} ) will be used as 
the muon trigger detector, in the barrel region of the ATLAS experiment at LHC \cite{Atlas}.
Good high $p_{t}$ muon trigger performance is crucial in order to
address the broad physics program of LHC.\\
In order to cover the barrel region a total number of 1116 RPC units will be installed, 
for a total surface area of about 3800 $m^{2}$.
The extreme difficulty in accessing the
ATLAS detectors, after installation is complete, imposes a high standard Quality Assurance
for these units.
For this purpose three cosmic ray teststands have been built at INFN Napoli\cite{Alviggi}, Lecce, and Roma 2,  each one allowing to certify 
a tower of eight RPC units at once. 
The Lecce site uses a standalone cosmic ray trigger and tracking system 
built with 4 pre-tested ATLAS units ( Figure \ref{fig1.eps} ). 
The advantages of this setup is a complete coverage of the chambers under test, avoiding time
consuming position scans, allowing only one readout system, and to monitor RPC behaviors in the long term
for the trigger chambers.
\section{Experimental setup and software components}
Our apparatus consists of several subsystems: gas distribution, high voltage distribution, low voltage distribution,
trigger logic, VME readout, and data acquisition. 
A large number of independent gas lines (one for each gas volume layer), high voltage and low voltage channels 
(one for each readout panel) are present, in order to easily isolated defective detectors.\\ 
The cosmic rays are selected with a loose trigger request and further refined on-line
requiring a minimum number of hits in the trigger RPC, resulting in an acquisition rate of about 50 Hz. 
Events are than analized looking for straight tracks (85\% of the events contain good track candidates). 
Those tracks are then used for 
efficiency computation of the RPCs under test and monitoring purposes.\\  
The monitor, control, and readout sofware is based on Labview, while, the analysis software is written in object
oriented C++ language. 
In addition, configurations, runs and results are managed by MySQL databases and presented via web interface.  
Finally, the data are displayed by PAW and ROOT macro's.
\section{Quality control tests}
The certification of the chambers is accomplished by a sequence of several measurements: 
a gas volume leak test, a HV ramp-up and ramp-down current curves, efficiency versus high voltage 
(at different front-end voltage threshold values),
single rate counts versus front-end voltage thresholds (at different high voltage values), and finally, chamber
tomography. The whole procedure (both for on-line and off-line DAQ processing) is almost automatic, 
takes about 24 hours, and starts after 2-days of gas flowing 
(94.7\%C$_2$H$_2$F$_4$-5\%C$_4$H$_{10}$-0.3\%SF$_6$).\\ 
\begin{figure}[htbp]
  \centerline{\hbox{\includegraphics[width=10cm]{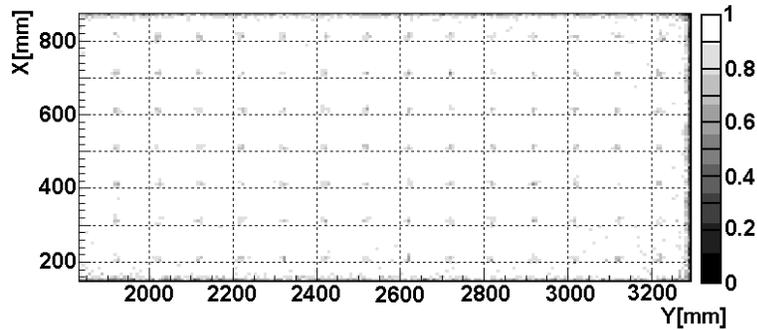}}
  }
 \caption{\it Tomography of a ATLAS RPC gas volume.\label{fig2.eps} }
\end{figure}  
During the gas volume leak test the differential pressure is monitored for two hours after imposing about 3 mbar of
over-pressure and closing the gas inlet and oulet. The runs with cosmic ray triggers (plateau curves and tomography plot) 
and with random trigger (noise rates) allow a complete characterization of the chambers. 
Up to now we tested about 100 units. 
Accepted chambers should have high gap efficiency ($>$ 97\% where about 2\% loss can be accounted due to the spacers) 
and relatively low single rate counts (about 1 $\frac{Hz}{cm^2}$, much less than the expected ATLAS cavern background).
Figure \ref{fig2.eps} shows a gas volume RPC tomography obtained with our system, proving the good 3D tracking capability 
of ATLAS RPC's (about 7 mm projected track spatial resolution in both projections). 
In fact, localized dead regions, due to the 1 cm diameter spacers, are clearly visible.
Moreover, the current versus high voltage allows to detect defective gap, 
looking for large leakage current.

\section{Conclusions}
The cosmic ray teststand at INFN Lecce is now capable of routinely testing ATLAS RPC detector units.  
Its complexity is such that the system is a good test bench for operating with a large number of RPC's.

\end{document}